\documentclass[ste,twoside]{stefano}

\usepackage{amsmath}
\usepackage{amssymb}
\usepackage{graphics}
\usepackage{rotating}

\def\makeheadbox{{%
\hbox to0pt{\vbox{\baselineskip=10dd\hrule\hbox
to\hsize{\vrule\kern3pt\vbox{\kern3pt
\hbox{  {\sf Foundations of Physics Letters} {\bf 14}, 37--50 (2001)}
\hbox{  {\sf hep-th/0103129} \hspace*{11.7cm} 
$\boldsymbol{\Sigma \delta \Lambda}$ }
\kern3pt}\hfil\kern3pt\vrule}\hrule}%
\hss}}}

\def\u{\leavevmode\hbox{\normalsize1\kern-4.2pt\large1}}
\def\f{\mbox{\tiny $\frac{1}{2}$}}
\def\s{\mbox{$\frac{1}{2}$}}

\def\mi{\mbox{\tiny $-$}}
\def\+{\mbox{\tiny $+$}}
\def\0{\mbox{\tiny $0$}}
\def\1{\mbox{\tiny $1$}}
\def\2{\mbox{\tiny $2$}}
\def\3{\mbox{\tiny $3$}}
\def\4{\mbox{\tiny $4$}}
\def\5{\mbox{\tiny $5$}}
\def\x{\mbox{\tiny $x$}}
\def\y{\mbox{\tiny $y$}}
\def\z{\mbox{\tiny $z$}}
\def\m{\mbox{\tiny $m$}}
\def\t{\mbox{\tiny $t$}}
\def\i{\mbox{\tiny $i$}}
\def\j{\mbox{\tiny $j$}}
\def\k{\mbox{\tiny $k$}}
\def\lep{\mbox{\tiny $l$}}
\def\qua{\mbox{\tiny $q$}}

\def\P{\mbox{\tiny $Par$}}
\def\Lo{\mbox{\tiny $Lor$}}
\def\L{\mbox{\tiny $L$}}
\def\R{\mbox{\tiny $R$}}
\def\F{\mbox{\tiny $F$}}
\def\Y{\mbox{\tiny $Y$}}
\def\D{\mbox{\tiny $D$}}
\def\C{\mbox{\tiny $C$}}
\def\M{\mbox{\tiny $M$}}

\def\sv{\mbox{$\frac{\varphi}{2}$}}
\def\sdag{\mbox{\tiny $\dag$}}

\def\tpm{\mbox{\tiny $\pm$}}

\def\q{\mbox{\tiny $(p)$}}
\def\e{\mbox{\tiny $\eta$}}

\begin{document}


\title{Quaternionic Lorentz group and Dirac equation}


\author{Stefano De Leo\inst{1,2} 
\thanks{Partially supported by the FAPESP grant 99/09008--5.}
}

\institute{
Department of Applied Mathematics, University of Campinas\\ 
PO Box 6065, SP 13083-970, Campinas, Brazil\\
{\em deleo@ime.unicamp.br}
\and
Department of Physics and INFN, University of Lecce\\ 
PO Box 193, I 73100, Lecce, Italy\\
{\em deleos@le.infn.it}
}


\date{Submitted {\em August 8, 2000} - Accepted {\em November 23, 2000}}

\abstract{
We formulate Lorentz group representations in which ordinary complex numbers 
are replaced by linear functions of real quaternions and  introduce dotted 
and undotted quaternionic one-dimensional spinors. To extend to parity
the space-time transformations, we combine  these one-dimensional spinors 
into bi-dimensional column vectors. 
From the transformation properties of the two-component  
spinors, we derive a quaternionic chiral representation for the 
space-time algebra. Finally, we obtain a quaternionic 
bi-dimensional version of the Dirac equation.
}




\maketitle

%
%


\section{Introduction}
\label{s1}

The isomorphism between unitary quaternions and space rotations is extended
to Lorentz boosts by means of 
linear functions of real quaternions. The use of 
linear functions of real quaternions in formulating Lorentz group 
representations lead to the 
introduction of two {\em inequivalent} one-dimensional spinors. 
In order to include parity, we construct bi-dimensional column 
vectors  which satisfy  a quaternionic version of the Dirac equation.  
For  massless particles, this equation 
decouples into one-dimensional equations, 
which represent the quaternionic counterparts of the Weyl equations. 
This formulation
of the Dirac equation allows to find a quaternionic chiral 
representation for the gamma matrices. This space-time matrix  representation 
can be used in formulating quaternionic gauge models for the 
electroweak interaction.  The use of linear functions of real quaternions  
and the  assumption of a complex  projection for the inner 
product~\cite{HOR,REM}, play a fundamental role to recover the mathematical 
properties of the Lorentz  group~\cite{DeSR}, to define an appropriate  
momentum operator~\cite{ROT}
and to obtain the right orthogonality between the solutions  of the Dirac 
equation~\cite{DeRod}.  

The paper is structured as follows. In section~\ref{s2}, we introduce
the quaternionic Lorentz group~\cite{DeSR} by using left/right 
operators~\cite{DIX,DeDu}.
Section~\ref{s3} is intended to study quaternionic dotted and undotted 
spinors. In such a section, we give the quaternionic chiral representation 
for the gamma matrices. 
In section~\ref{s4}, we   
discuss the quaternionic Dirac 
equation~\cite{ROT,MOR,DeDIR,DeRodDIR,DeRoVa} 
and justify the adoption of a complex projection 
for the inner product. In section~\ref{s5}, we give a brief introduction to
quaternionic gauge models. Conclusions and out-looks are drawn in the final
section.


\section{Linear functions of real quaternions and Lorentz group}
\label{s2}

A generic quaternion $q$ can be defined in terms of
real~\cite{HAM} or complex~\cite{GUR} numbers as follows  
\begin{equation}
q = a_{\0}+ia_{\1}+ja_{\2}+ka_{\3} = \xi + j \eta~,~~~~~
a_{\0 , \1 , \2 , \3 } \in \mathbb{R}~,~~~\xi , \eta \in \mathbb{C}(1,i)~,
\end{equation}
where the imaginary units $i$, $j$ and $k$ satisfy
\begin{equation}
i^{\2}=j^{\2}=k^{\2}=ijk= - 1~.
\end{equation}
The conjugate of $q$, denoted by $\bar{q}$, is given by 
$
\bar{q}  = a_{\0}-ia_{\1}-ja_{\2}-ka_{\3} = \xi^* - j \eta~.
$

Due to the non-commutative nature of the quaternionic field,  
we have to  distinguish between the {\em left} and 
{\em right} action of the imaginary units $i$, $j$, $k$. 
To do it, we introduce linear functions of real quaternions.
For example, we can use the linear function $i \, (~~) \, j$ to 
represent the simultaneous action
on a quaternionic state $q$ of the imaginary unit $i$ from the left
and of the imaginary unit $j$ from the right, i.e. $i \, q \, j$.
In order to shorten our notation, we shall rewrite linear functions of 
real quaternions in terms of the  left/right operators~\cite{DIX,DeDu}
\begin{equation}
\boldsymbol{L} \equiv \left( L_{\i} , L_{\j} , L_{\k} \right)~~~~\mbox{and}~~~~
\boldsymbol{R} \equiv \left( R_{\i} , R_{\j} , R_{\k} \right)~,
\end{equation}
\[
\boldsymbol{L}~:~\mathbb{H} \rightarrow \mathbb{H}~,~~~
\boldsymbol{L} \, q \equiv \boldsymbol{h} \, q~~~~\mbox{and}~~~~
\boldsymbol{R}~:~\mathbb{H} \rightarrow \mathbb{H}~,~~~
\boldsymbol{R} \, q \equiv q \, \boldsymbol{h}~,~~~~~
\boldsymbol{h} \equiv \left( i,j,k \right)~.
\]
In this formalism, the linear function of real quaternions $i \, (~~) \, j$
is concisely expressed by $L_{\i} R_{\j}$.  The algebra of left and right 
operators is given by  
\[
L_{\i}^{\2} = L_{\j}^{\2} =L_{\k}^{\2} = L_{\i} L_{\j} L_{\k} = 
R_{\i}^{\2} = R_{\j}^{\2} =R_{\k}^{\2} = R_{\k} R_{\j} R_{\i} = - 
\boldsymbol{1}~,
\]
and by the commutation relations
\[
\left[ \, L_{i,j,k} ~ , ~ R_{i,j,k} \, \right] = 0~.
\]

The Lorentz group $\mathsf{SO}(3,1)$ is  characterized by six generators. 
The anti-hermitian generators associated to the space rotations, 
$\boldsymbol{\mathcal{A}}$, and the hermitian boost  generators, 
$\boldsymbol{\mathcal{B}}$, satisfy the following commutation relations
\begin{eqnarray}
\label{com2}
\mathcal{A}_{\x} = [ \mathcal{A}_{\y} , \mathcal{A}_{\z} ]~,~~~ 
\mathcal{A}_{\x} = [ \mathcal{B}_{\z} , \mathcal{B}_{\y} ]~,~~~
\mathcal{B}_{\x} = [ \mathcal{A}_{\y} , \mathcal{B}_{\z} ]
              = [ \mathcal{B}_{\y} , \mathcal{A}_{\z} ]~, \nonumber\\
\mathcal{A}_{\y} = [ \mathcal{A}_{\z} , \mathcal{A}_{\x} ]~,~~~ 
\mathcal{A}_{\y} = [ \mathcal{B}_{\x} , \mathcal{B}_{\z} ]~,~~~ 
\mathcal{B}_{\y} = [ \mathcal{A}_{\z} , \mathcal{B}_{\x} ]
              = [ \mathcal{B}_{\z} , \mathcal{A}_{\x} ] ~,\\ 
\mathcal{A}_{\z} = [ \mathcal{A}_{\x} , \mathcal{A}_{\y} ]~,~~~
\mathcal{A}_{\z} = [ \mathcal{B}_{\y} , \mathcal{B}_{\x} ]~,~~~ 
\mathcal{B}_{\z} = [ \mathcal{A}_{\x} , \mathcal{B}_{\y} ]
              = [ \mathcal{B}_{\x} , \mathcal{A}_{\y} ]~. \, \nonumber
\end{eqnarray}
By using the left quaternionic imaginary units $\boldsymbol{L}$ and the 
right complex imaginary unit $R_{i}$, 
 we can obtain a one-dimensional representation for the
Lorentz  generators~\cite{DeSR}. The rotation generators  are given in terms 
of the left-acting quaternionic imaginary units,  
\begin{equation}
\label{ql1}
\mathcal{A}_{\x} = L_{\i}/2~,~~~
\mathcal{A}_{\y} = L_{\j}/2~,~~~
\mathcal{A}_{\z} = L_{\k}/2~~~~~\in \mathbb{H}^{\L}~.
\end{equation}
The boost generators are represented by the joint action of the  
left-quaternionic imaginary units, $i$, $j$ and $k$,  and of the right-complex 
imaginary unit $i$
\begin{equation}
\label{ql2}
\mathcal{B}_{\x} = \pm L_{\i} R_{\i} /2~,~~~
\mathcal{B}_{\y} = \pm L_{\j} R_{\i}/2~,~~~
\mathcal{B}_{\z} = \pm L_{\k} R_{\i}/2~~~~~\in 
\mathbb{H}^{\L} \otimes \mathbb{C}^{\R}~.
\end{equation}
These two {\em inequivalent} quaternionic representations for the 
boost generators imply two different transformation laws for 
quaternionic 
spinors,  
\begin{equation}
s_{\+} = \exp \left[ \boldsymbol{L} \cdot 
                         \left( \boldsymbol{\theta} + R_{\i} \, \,
                    \boldsymbol{\varphi} \right) /2
                       \right]~~~~\mbox{and}~~~~
s_{\mi} = \exp \left[ \boldsymbol{L} \cdot 
                         \left( \boldsymbol{\theta} - R_{\i} \, \,
                    \boldsymbol{\varphi} \right) /2
                  \right]~,
\end{equation}
where 
$\boldsymbol{\theta}   \equiv 
\left( \theta_{\x} , \theta_{\y} , \theta_{\z} \right)$ and 
$\boldsymbol{\varphi} \equiv \left( \varphi_{\x} , \varphi_{\y} , 
\varphi_{\z} \right)$. 
For consistence, we introduce two 
quaternionic spinors, $q_{\tpm}= \xi_{\tpm} + j \eta_{\tpm}$,
which, under the action of the quaternionic unitary group,   
transform into $s_{\tpm} \, q_{\tpm}$.

For a detailed review of quaternionic group theory, 
we refer the reader to the 
Gilmore book~\cite{GIL} and  refs.~\cite{DeDu,ADL,DeGUT,Sco1,DeSco,Sco2}.


\section{Dotted and undotted spinors}
\label{s3}

As remarked in the previous section, the two 
possible signs of $\boldsymbol{{\cal B}}$ in equation(\ref{ql2}) imply
\[
q_{\+} ~\rightarrow~ {s_{\+}} \, q_{\+} = \exp \left[ \boldsymbol{L} \cdot 
                      \left( \boldsymbol{\theta} + R_{\i} \, \,
                      \boldsymbol{\varphi} \right) /2
                       \right] \, q_{\+}~~~\mbox{and}~~~
q_{\mi} ~\rightarrow~
                      {s_{\mi}} \, q_{\mi} = \exp \left[ \boldsymbol{L} \cdot 
                       \left( \boldsymbol{\theta} - R_{\i} \, \,
                       \boldsymbol{\varphi} \right) /2
                       \right] \, q_{\mi}~.
\]
These {\em inequivalent} representations 
of the Lorentz group are characterized by two
different types of one-component quaternionic spinors, $q_{\tpm}$,
which represent the quaternionic  counterpart of the standard 
{\em dotted} and {\em undotted} bi-dimensional complex spinors. 
These spinors correspond to the 
representations
$\left( \s , 0 \right)$ and $\left( 0 , \s \right)$ 
of the Lorentz group. The generators
\begin{equation}
\boldsymbol{{\cal A}}_{\1} = \s \left( 
                     \boldsymbol{{\cal A}} - \boldsymbol{{\cal B}} \, R_{\i} \right)
~~~\mbox{and}~~~
\boldsymbol{{\cal A}}_{\2} = \s \left(
                      \boldsymbol{{\cal A}} + \boldsymbol{{\cal B}} \, R_{\i} \right)~,
\end{equation}
satisfy the following commutation relations
\[
\mathcal{A}_{\m , x} = 
[ \, \mathcal{A}_{\m , y} \, , \, \mathcal{A}_{\m , z} \,]
~,~~~ 
\mathcal{A}_{\m , y} = 
[ \,  \mathcal{A}_{\m , z} \,  , \,   \mathcal{A}_{\m , x} \,  ]
~,~~~ 
\mathcal{A}_{\m , z} = 
[ \,  \mathcal{A}_{\m , x} \, , \, \mathcal{A}_{\m , y} \, ]~,~~~~~
\mbox{\tiny $m = 1,2$}~,
\]
and
\[ 
\left[ \, \boldsymbol{\mathcal{A}}_{\1} \, , \, 
\boldsymbol{\mathcal{A}}_{\2} \, \right] = 0~.
\]  
Thus, the quaternionic  representations of the Lorentz group can be 
classified  in terms of the representations of the left-acting group 
group $\mathsf{U}_{\1}(1,\mathbb{H}^{\L}) \otimes 
\mathsf{U}_{\2}(1,\mathbb{H}^{\L})$.
Quaternionic  states will be identified by two angular momenta 
$\left( j_{\1} , j_{\2} \right)$ corresponding to 
the generators $\boldsymbol{\cal A}_{\1}$ and 
$\boldsymbol{\cal A}_{\2}$. In our case, 
\begin{eqnarray*}
   \boldsymbol{\cal A} = \boldsymbol{L} /2~,~~~ \boldsymbol{\cal B} = + \boldsymbol{L} R_{\i} /2
   ~~~\Rightarrow~~~
\boldsymbol{\cal A}_{\1} = \boldsymbol{L} /2~,~~~   
\boldsymbol{\cal A}_{\2} = 0~,~~~&     \left( \s , 0 \right)~,\\
   \boldsymbol{\cal A} = \boldsymbol{L} /2~,~~~ \boldsymbol{\cal B} = - \boldsymbol{L} R_{\i} /2
   ~~~\Rightarrow~~~
\boldsymbol{\cal A}_{\2} = \boldsymbol{L} /2~,~~~   
\boldsymbol{\cal A}_{\1} = 0~,~~~&     \left( 0 , \s \right)~.
\end{eqnarray*}
Under parity, velocity changes sign, 
hence the boost generators change sign,
$\boldsymbol{{\cal B}} \rightarrow - \boldsymbol{{\cal B}}$. 
The rotation generators behave like axial vectors, 
$\boldsymbol{{\cal A}} \rightarrow \boldsymbol{{\cal A}}$. 
So, space inversion imply 
$(j_{\1},0) \leftrightarrow  (0,j_{\2})$ and consequently 
$q_{\+} \leftrightarrow q_{\mi}$. It is no longer 
sufficient to consider one-dimensional 
quaternionic spinors. We need to combine $q_{\+}$ and $q_{\mi}$ into
bi-dimensional quaternionic column vectors
\begin{equation}
\psi = \left( \begin{array}{c} q_{\+} \\q_{\mi} \end{array} \right)~.
\end{equation}
The bi-dimensional spinor $\psi$ is an {\em irreducible} 
representation of the Lorentz group {\em extended} by parity,
\[
\psi \rightarrow  \left( \begin{array}{ll} s_{\+}  & 0\\
                                            0 &  s_{\mi}           
                  \end{array} \right)_{\Lo}  
      \left( \begin{array}{c} q_{\+} \\q_{\mi} \end{array} \right)
~~~\mbox{and}~~~
\psi \rightarrow  \left( \begin{array}{cc} \, 0 \,  & \, 1 \, \\
                                           1  & 0           
                  \end{array} \right)_{\P}  
      \left( \begin{array}{c} q_{\+} \\q_{\mi} \end{array} \right)~.
\]
Under Lorentz boosts, 
\[
q_{\+} \rightarrow \exp \left( \boldsymbol{L} R_{\i} \cdot \boldsymbol{\varphi}  / 2
                       \right) \, q_{\+} = 
                   \exp \left( L_{\boldsymbol{n}} R_{\i} \varphi  / 2
                       \right) \, q_{\+} =
\left( \cosh \sv + L_{\boldsymbol{n}} R_{\i} \sinh \sv \right) 
                   q_{\+}~,
\]
where
\[ 
L_{\boldsymbol{n}} \equiv \boldsymbol{L} \cdot \boldsymbol{n} = 
n_{\1} L_{\i} + n_{\2} L_{\j} + n_{\3} L_{\k}~,~~~
L_{\boldsymbol{n}}^{\2} = - \mathbf{1}~,~~~
\mbox{\small 
$\varphi = \sqrt{ \varphi_{\x}^{\2} + \varphi_{\y}^{\2} + 
\varphi_{\z}^{\2}}$}~.
\]
The $q_{\mi}$ transformation is soon obtained from the previous one
by changing $\varphi \rightarrow - \varphi$,
\[
q_{\mi} \rightarrow \left( \cosh \sv - L_{\boldsymbol{n}} R_{\i} \sinh \sv \right) 
                   q_{\mi}~.
\]
We can identify the original spinors with the  particle at rest, 
$q_{\tpm} (0)$,
and the transformed spinors, $q_{\tpm} (\boldsymbol{p})$, with the particle 
moving with momentum $\boldsymbol{p}$. By observing that
\[
\cosh \varphi = \gamma = E/m~,~~~\sinh \varphi = \beta \gamma = p / m~,
~~~~~(c=1)~,
\]
and
\[ 
\cosh \sv = \left( \frac{\gamma + 1}{2} \right)^{\f} =
            \left( \frac{E+ m}{2m} \right)^{\f} ~,~~~
\sinh \sv = \left( \frac{\gamma - 1}{2} \right)^{\f} =
            \left( \frac{E- m}{2m} \right)^{\f}~,
\]
we find
\begin{eqnarray*}
q_{\+} (\boldsymbol{p})  =  \left[  \left( \frac{E+ m}{2m} \right)^{\f} + 
                             L_{\boldsymbol{n}} R_{\i}             
            \left( \frac{E- m}{2m} \right)^{\f} \right] \, q_{\+} (0)
 =   \left( E + m + \boldsymbol{L} R_{\i} \cdot \boldsymbol{p} \right)
                   \left[ 2m \left( E+m \right) \right]^{\mi \f} \, q_{\+} 
(0)~ \, 
\end{eqnarray*}
and
\begin{eqnarray*} 
q_{\mi} (\boldsymbol{p}) =  \left[  \left( \frac{E+ m}{2m} \right)^{\f} - 
                             L_{\boldsymbol{n}} R_{\i}             
            \left( \frac{E- m}{2m} \right)^{\f} \right] \, q_{\mi} (0)
             =  \left( E + m - \boldsymbol{L} R_{\i} \cdot 
\boldsymbol{p} \right)
                  \left[ 2m \left( E+m \right) \right]^{\mi \f} \, 
q_{\mi} (0)~.
\end{eqnarray*} 
By recalling that $(0)$ corresponds to the rest frame frame of the particle, 
we set $q_{\+} (0) = q_{\mi} (0)$. So,  we obtain
\[
m \, q_{\+} (\boldsymbol{p})  = 
\left( E + \boldsymbol{L} R_{\i} \cdot \boldsymbol{p}  \right) \, 
q_{\mi} (\boldsymbol{p}) ~~~\mbox{and}~~~
m \, q_{\mi} (\boldsymbol{p}) = 
\left( E - \boldsymbol{L} R_{\i} \cdot \boldsymbol{p}  \right) \, q_{\+} 
(\boldsymbol{p})~.
\]
These equations can be rewritten in matrix form as 
\begin{equation}
\label{de}
\left( \begin{array}{cc} $-$ m & E + \boldsymbol{L} R_{\i} \cdot 
\boldsymbol{p} \\
         E - \boldsymbol{L} R_{\i}  \cdot \boldsymbol{p}  & $-$ m \end{array}
\right)
\left[ \begin{array}{c} q_{\+} (\boldsymbol{p})  \\
 q_{\mi} (\boldsymbol{p}) \end{array} \right] = 0~.
\end{equation}
By introducing the following complex linear quaternionic representation for 
the  gamma matrices
\begin{equation}
\label{chi}
\gamma^{\0} \equiv \left( \begin{array}{cc} \, 0 \, & \, 1 \, \\
                         1 & 0 \end{array} \right)~,~~~
\boldsymbol{\gamma} \equiv \left( \begin{array}{rr} \, 0 \, & $-$  1 \\
                          1 \,  & 0 \end{array} \,
\right) \, \boldsymbol{L}  R_{\i}   ~,
\end{equation}
we find
\begin{equation}
\label{dirac}
( \gamma^{\mu} p_{\mu} - m ) \psi (\boldsymbol{p}) = 0~.
\end{equation}
For massless particles, 
this  equation decouples into one-dimensional equations,
\[
\left( E + \boldsymbol{L} R_{\i} \cdot \boldsymbol{p} \right) \, 
q_{\mi} (\boldsymbol{p}) = 0~~~\mbox{and}~~~
\left( E - \boldsymbol{L} R_{\i} \cdot \boldsymbol{p}  \right) \, 
q_{\+} (\boldsymbol{p}) = 0~.
\]
Since,
for a massless particle $E=p$, these equations can also be rewritten as
\begin{equation}
\boldsymbol{L} R_{\i}  \cdot 
\hat{\boldsymbol{p}}  \, q_{\mi} (\boldsymbol{p}) = 
- q_{\mi} (\boldsymbol{p})~~~\mbox{and}~~~ 
\boldsymbol{L} R_{\i} \cdot 
\hat{\boldsymbol{p}}   \, q_{\+} (\boldsymbol{p})  
= q_{\+} (\boldsymbol{p})~. 
\end{equation}
The operator 
$\boldsymbol{L} R_{\i} \cdot \hat{\boldsymbol{p}}$ 
measures the component of the spin
in the direction of momentum and defines the 
{\em quaternionic helicity operator}.

\section{Chiral and Dirac representations}
\label{s4}

The derivation of equation (\ref{dirac}), which after quantization 
leads to a quaternionic version of the  Dirac equation, 
differs from the original one formulated by Rotelli~\cite{ROT}.
The equation presented in this paper is directly obtained 
from the transformation properties of the quaternionic Lorentz spinors, 
whereas the  Rotelli derivation follows the original Dirac approach. 
In the Rotelli formulation, we have the following representation for the 
gamma matrices
\[ 
\gamma^{\0}_{\D} \equiv \left( \begin{array}{rr} \, 1  \, &  0 \, \\
                         0 \,  & $-$ 1 \end{array} \right)~,~~~
\boldsymbol{\gamma}_{\D} \equiv \left( \begin{array}{cc} \, 0 \,  & \, 1 \, \\
                          1 & 0 \end{array}
\right) \, \boldsymbol{L}~.
\]
In this case,  the  space-time algebra $\mathsf{Cl}_{\1 , \3}$ is
given in terms of left-acting  operators, $\gamma^{\mu} \in 
\mathcal{M}_{\2}(\mathbb{H}^{\L})$.
In our formulation, we find a complex linear  set of gamma-matrices 
\[
\gamma^{\0}_{\C} \equiv \left( \begin{array}{cc} \, 0 \, & \, 1 \, \\
                         1 & 0 \end{array} \right)~,~~~
\boldsymbol{\gamma}_{\C} \equiv \left( \begin{array}{rr} \, 0 \, & $-$  1 \,\\
                          1 \,  & 0 \,  \end{array}
\right) \, \boldsymbol{L} R_{\i}~.
\]
Thus, we get a $\mathbb{C}$-linear quaternionic space-time algebra,
$\gamma^{\mu} \in \mathcal{M}_{\2}(\mathbb{H}^{\L} \otimes \mathbb{C}^{\R})$.
We observe that the $\mathbb{C}$-linearity is only hidden in the 
Rotelli approach. In the complex world, the Dirac equation 
reads indifferently as
\[ 
i \partial_t \psi = H \psi ~~~\mbox{or}~~~
   \partial_t \psi i = H \psi ~.
\]
In the quaternionic world there is a clear difference in choosing a left or 
right position for the complex imaginary unit $i$. In fact, 
by requiring norm conservation
\[ \partial_t \int d^{\3} x \, \psi^{\sdag} \psi  = 0~, \]
we find that a left position of the imaginary unit $i$ in the quaternionic
Dirac equation,  that is 
\[ L_{\i} \partial_t \psi \equiv i \partial_t \psi = H \psi~,  \]
implies
\[ 
\partial_t \int d^{\3} x \, \psi^{\sdag} \psi  = 
\int d^{\3} x \, \psi^{\sdag} [ H , i ] \psi
\]
in general $\neq 0$ for a quaternionic Hamiltonian. A right position
of the imaginary unit $i$, 
\[ R_{\i} \partial_t \psi \equiv \partial_t \psi i = H \psi~,  \]
ensures the norm conservation. By treating time and space in 
the same way, we gain the following definition for the energy/momentum 
operator 
\begin{equation}
p^{\mu} \leftrightarrow R_{\i} \partial^{\mu}~~~\Rightarrow~~~
p^{\mu} \psi \leftrightarrow R_{\i} \partial^{\mu} \psi \equiv 
\partial^{\mu} \psi i~.
\end{equation}
Finally, the complex linear quaternionic Dirac equation reads
\begin{equation}
\label{dirac1}
R_{\i} \gamma^{\mu} \partial_{\mu} \psi \equiv 
\gamma^{\mu} \partial_{\mu} \psi \, i = m \psi~,~~~
\left[ \gamma^{\mu} ,  \gamma^{\nu} \right] = 2 g^{\mu \nu}~.
\end{equation}
Under Lorentz transformations, we get
\[ 
\gamma^{\mu}  a_{\mu \nu} \partial^{\nu} \mathcal{S} \, \psi \, i = m  
\, \mathcal{S} \, \psi~,~~~a_{\mu \nu} \in \mathbb{R}~.
\]
In order to guarantee the covariance of the Dirac equation, we must have
\[
a_{\mu \nu} \, \mathcal{S}^{\mi \1} \, \gamma^{\mu} \, \mathcal{S} =
\gamma_{\nu}~.
\]
For infinitesimal transformations, 
$a_{\mu \nu} = g_{\mu \nu} + \omega_{\mu \nu}$, we write
\[
\mathcal{S} = \boldsymbol{1} - \mbox{$\frac{1}{4}$} \, \sigma_{\mu \nu} \,
\omega^{\mu \nu} \, R_{\i}~.
\]
A set of matrices $\sigma_{\mu \nu}$ satisfying this relation is given by
$\frac{1}{2} \, \left[ \gamma_{\mu} , \gamma_{\nu} \right] \, R_{\i}$. 
A finite transformation is then given by
\[
\mathcal{S} = 
\exp \left( \mbox{$\frac{1}{8}$} \, 
\left[ \, \gamma^{\mu} ,  \gamma^{\nu} \right] \,
\omega_{\mu \nu} 
\right)~.
\]

A fundamental ingredient in the formulation of quaternionic relativistic 
quantum mechanics is represented by the adoption of a 
{\em complex projection} of the inner product~\cite{HOR,ROT},
necessary in order to guarantee  that 
$R_{\i} \boldsymbol{\partial}$ be an hermitian operator. In fact, 
\[ 
\int d^{\3} x \, \varphi^{\sdag} R_{\i} \boldsymbol{\partial} \psi  = 
\int d^{\3} x \,  ( R_{\i} \boldsymbol{\partial} \varphi ) ^{\sdag}  
\psi ~~~\Rightarrow~~~
\int d^{\3} x \, \varphi^{\sdag} \boldsymbol{\partial} \psi \, i  = 
- i \int d^{\3} x \,  \boldsymbol{\partial} \varphi^{\sdag}  \psi~. 
\]
After integration by parts in the last integral, we find 
\[ 
\int d^{\3} x \, \varphi^{\sdag} \boldsymbol{\partial} \psi \, i  = 
 i \int d^{\3} x \, \varphi^{\sdag} \boldsymbol{\partial} \psi~.
\]
Due to the different position of the imaginary unit $i$ and
to the quaternionic nature of the wave functions, the previous relation
is only  satisfied by adopting a 
complex projection for the inner product,
\begin{equation}
\int d^{\3} x  ~\rightarrow~ \int_{\mathbb{C}} d^{\3} x~.  
\end{equation}
The use of a complex projection of the inner product also gives 
the right number of  
orthogonal solutions for the quaternionic 
Dirac equation~\cite{ROT}.

\section{Quaternionic electroweak gauge models}
\label{s5}

Let us rewrite equation (\ref{dirac}) as
\begin{equation}
\label{dirac2}
\Upsilon^{\mu} \partial_{\mu} \psi = m \psi~,~~~
\Upsilon^{\mu} \equiv \gamma^{\mu} R_{\i}~,~~~
\left[ \Upsilon^{\mu} ,  \Upsilon^{\nu} \right] = 
- 2 g^{\mu \nu}~.
\end{equation}
By using the quaternionic Dirac and chiral representations
for the gamma matrices, we find
\[ 
\Upsilon^{\0}_{\D} \equiv \left( \begin{array}{rr} \, 1 \, & 0 \\
                         0 \, & $-$ 1 \end{array} \right) \, R_{\i} ~,~~~
\boldsymbol{\Upsilon}_{\D} \equiv \left( \begin{array}{cc} \, 0 \, & \, 1 \,\\
                          1 & 0 \end{array}
\right) \, \boldsymbol{L} R_{\i}~~~\mbox{and}~~~
\Upsilon^{\0}_{\C} \equiv \left( \begin{array}{cc} \, 0 \, & \, 1 \,\\
                         1 & 0 \end{array} \right) \, R_{\i} ~,~~~
\boldsymbol{\Upsilon}_{\C} \equiv \left( \begin{array}{rr} 0 & \, 1 \, \\
                          $-$ 1 & 0 \, \end{array}
\right) \, \boldsymbol{L} ~.
\]
From the quaternionic space-time algebra,  we can  extract the generators
of a right-acting complex $\mathsf{SU}(2)$ group. In fact, the matrices
$\Upsilon^{\0}$, 
$\Upsilon^{\1 \2 \3}= \Upsilon^{\1} \Upsilon^{\2} \Upsilon^{\3}$ and 
$\Upsilon^{\5} = \Upsilon^{\0} \Upsilon^{\1 \2 \3}$ satisfy the 
following algebra
\[
\left( \Upsilon^{\0} \right)^{\2}=
\left( \Upsilon^{\1 \2 \3} \right)^{\2}=
\left( \Upsilon^{\5} \right)^{\2}=
\Upsilon^{\0} \Upsilon^{\1 \2 \3} \Upsilon^{\5}= - \mathbf{1}
\]
and represent right-acting complex operators
\begin{equation*}
\left( \, \Upsilon^{\0}_{\D}~,~
\Upsilon^{\1 \2 \3}_{\D} ~,~
\Upsilon^{\5}_{\D} \, 
\right) 
\equiv 
\left[ \, \left( \begin{array}{rr} \, 1 \, & 0 \\
                         0 \, & $-$ 1 \end{array} \right) \, R_{\i}~,~
 \left( \begin{array}{cc} \, 0 \, & \, 1 \,  \\
                         1  & 0 \end{array} \right) \, R_{\i}~,~
 \left( \begin{array}{rr} \, 0 \, & $-$ 1 \\
                         1 \, & 0 \end{array} \right) \, 
\right]~ \,   
\end{equation*}
and
\begin{equation*}
\left( \, \Upsilon^{\0}_{\C}~,~
\Upsilon^{\1 \2 \3}_{\C}~,~
\Upsilon^{\5}_{\C} \, 
\right) 
\equiv 
\left[ \,  \left( \begin{array}{cc} \, 0 \, & \, 1 \,  \\
                         1  & 0 \end{array} \right) \, R_{\i}~,~
 \left( \begin{array}{rr}  0 \, &  \, 1 \,  \\
                         $-$ 1 \, & 0 \, \end{array} \right) ~,~
\left( \begin{array}{rr} $-$ 1 \, & 0 \, \\
                         0 \, & 1 \, \end{array} \right) \, R_{\i} \,
\right]~.  
\end{equation*}
The Dirac and chiral representations diagonalize respectively  
$\Upsilon^{\0}$ and $\Upsilon^{\5}$. By diagonalizing 
$\Upsilon^{\1 \2 \3}$, we gain a {\em new} set of gamma matrices
\[ 
\Upsilon^{\0}_{\M} \equiv \left( \begin{array}{cc} \, 0 \, & \, 1 \, \\
                         1 & 0 \end{array} \right) \, R_{\i} ~,~~~
\boldsymbol{\Upsilon}_{\M} \equiv \left( \begin{array}{rr} \, 1 \, & 0 \, \\
                          0 \, & $-$ 1 \, \end{array}
\right) \,  \boldsymbol{L} R_{\i}~,
\]
quaternionic translation of the Majorana representation.
Finally, the Dirac, Majorana and chiral representations can be associated
to the three possible diagonalization choices for the generators of 
the right-acting 
complex group $\mathsf{SU}(2, \mathbb{C}^{\R})$, i.e. 
$\Upsilon^{\0}$, $\Upsilon^{\1 \2 \3}$ and  $\Upsilon^{\5}$.

We can write the first fermion family of the  Salam-Weinberg 
model
by {\small $2 \times 2$} quaternionic  
matrices, $\Psi_{\lep}$ 
($\nu$-$e$ leptons) and $\Psi_{\qua}$ ($u$-$d$ quarks).
The complex projection of  the massless fermion electroweak 
Lagrangian~\cite{LAG} 
\begin{equation}
\label{lag}
\mathcal{L}_{\F} = \left[   
\bar{\Psi}_{\lep} \Upsilon^{\mu} \partial_{\mu} \Psi_{\lep} +
\bar{\Psi}_{\qua} \Upsilon^{\mu} \partial_{\mu} \Psi_{\qua} 
\right]_{\mathbb{C}}
\end{equation}
is global invariant under the gauge group 
$\mathsf{SU}(2, \mathbb{C}^{\R}) \otimes \mathsf{U}(1, \mathbb{C}^{\R})$.
In the chiral representation, we have  
\begin{equation}
\Psi_{\lep} = \left( \begin{array}{cc} \nu_{\R} &  e_{\R}\\
                                \nu_{\L} & e_{\L} 
\end{array} \right)~~~\mbox{and}~~~
\Psi_{\qua} = \left( \begin{array}{cc} u_{\R} &  d_{\R}\\
                                u_{\L} & d_{\L} 
\end{array} \right)~.
\end{equation}
The Lagrangian (\ref{lag}), rewritten in terms of these spinors
and of the quaternionic differential operators  
$\mathcal{D}_{\tpm} = \partial_{\t} R_{\i}  \pm 
\boldsymbol{L} \cdot \boldsymbol{\partial}$, reads
\begin{eqnarray}
\label{lag2}
\mathcal{L}_{\F} & = & \left[   
 \left( \begin{array}{cc} \nu_{\R} &  e_{\R}\\
                                \nu_{\L} & e_{\L} 
\end{array} \right)^{\sdag} 
 \left( \begin{array}{cc} 
\mathcal{D}_{\mi} &  0\\
0                 & \mathcal{D}_{\+} 
\end{array} \right) 
\left( \begin{array}{cc} \nu_{\R} &  e_{\R}\\
                                \nu_{\L} & e_{\L} 
\end{array} \right)
 +
 \left( \begin{array}{cc} u_{\R} &  d_{\R}\\
                                u_{\L} & d_{\L} 
\end{array} \right)^{\sdag}
\left( \begin{array}{cc} 
\mathcal{D}_{\mi} &  0\\
0                 & \mathcal{D}_{\+} 
\end{array} \right)  
\left( \begin{array}{cc} u_{\R} &  d_{\R}\\
                                u_{\L} & d_{\L} 
\end{array} \right)
\right]_{\mathbb{C}} \nonumber \\
 & = &
\left[   
 \left( \begin{array}{cc} \nu_{\R} &  e_{\R}\\
                           u_{\R} & d_{\R} 
\end{array} \right)^{\sdag} 
\mathcal{D}_{\mi}   
\left( \begin{array}{cc} \nu_{\R} &  e_{\R}\\
                           u_{\R} & d_{\R} 
\end{array} \right)
 +
 \left( \begin{array}{cc} \nu_{\L} &  e_{\L}\\
                                u_{\L} & d_{\L} 
\end{array} \right)^{\sdag}
\mathcal{D}_{\+} 
\left( \begin{array}{cc} \nu_{\L} &  e_{\L}\\
                                u_{\L} & d_{\L} 
\end{array} \right)
\right]_{\mathbb{C}} \nonumber \\
  & = &
\left[   
 \Psi_{\R}^{\, \sdag} 
\mathcal{D}_{\mi}   
\Psi_{\R}
 +
 \Psi_{\L}^{\, \sdag} \mathcal{D}_{\+} 
\Psi_{\L}\right]_{\mathbb{C}}~.
\end{eqnarray}
In a forthcoming paper
will be discussed in detail a quaternionic electroweak
theory based on the Lagrangian (\ref{lag2}) and the right-acting 
complex gauge group 
\[
g_{\2} \, \mathsf{SU}(2, \mathbb{C}^{\R})_{\L} \otimes 
g_{\Y} \mathsf{U}(1, \mathbb{C}^{\R})_{\Y}~.
\]
Formulations of left/right symmetric models~\cite{LR} require the 
generalization to the gauge group
\[
g_{\2 , \L } \, \mathsf{SU}(2, \mathbb{C}^{\R})_{\L} \otimes 
g_{\2 , \R } \, \mathsf{SU}(2, \mathbb{C}^{\R})_{\R} \otimes 
g_{\1} \mathsf{U}(1, \mathbb{C}^{\R})~.
\]
Grand unification models and super-symmetric theories 
could require the choice of {\em effective} quaternionic gauge 
groups.

\section{Conclusions}
\label{s6}

This work was intended as an attempt to motivate the use of quaternions in 
physics. In the last years, it was investigated the possibility to 
formulate quantum mechanics and field theory by using quaternionic 
wave functions and complex inner products.  
The use of a complex projection of the inner 
product~\cite{HOR,REM} is a fundamental 
ingredient to obtain a consistent definition of energy/momentum 
operator~\cite{ROT}. Complex inner products also appear in the Dirac-Hestenes 
equation~\cite{DH1,DH2,DH3} and in many other mathematical and physical 
applications of Clifford algebras~\cite{HES}.  
 In the Hestenes work, the ordinary commutative $i$ of standard quantum  
mechanics acquires a geometrical 
interpretation as the generator of rotations in a space-like plane. The
electron spin and complex numbers are combined in a single geometric 
entity~\cite{DH1,DH2,DH3}. The Hestenes theory based on the Pauli algebra 
(even projection of the space-time algebra, STA) is algebraically isomorphic 
to Dirac theory. The Hestenes and Dirac theories are
algebraically isomorphic but {\em not} equivalent. In fact, 
in STA formulation of the Dirac equation, we can give  a 
geometric interpretation of the ordinary 
commutative imaginary  unit $i$. Such an interpretation 
is {\em hidden} in the standard approach.

In terms of {\em complexified} quaternions,  i.e. by  introducing in
$\mathbb{H}$ an imaginary unity $\eta$ which commutes with the quaternionic 
imaginary units  $i$, $j$ and $k$, the Dirac  spinor is represented by
\[ \Psi = \psi_{\1} + j \, \psi_{\2} + \eta \, \left( 
 \psi_{\3} + j \, \psi_{\4} \right)~,~~~\psi_{\1,...,\4} \in \mathbb{C}(1,i)~.
\]
In this formalism~\cite{DeRod}, the Dirac-Hestenes  equation reads 
\begin{equation}
\label{bv}
\left( \partial_{t} + \eta \, \boldsymbol{h} \cdot \boldsymbol{\partial} 
\right) \Psi \, i  = m \, \tilde{\Psi}~,
\end{equation}
where 
$\tilde{\Psi} = \psi_{\1} + j \, \psi_{\2} - \eta \, \left( 
 \psi_{\3} + j \, \psi_{\4} \right)$. 
In the rest frame, we find the following solutions
\[ \Psi_{\1} = \exp [i m \tau]~,~~
 \Psi_{\2} = j \, \exp [i m \tau]~,~~
 \Psi_{\3} =  \eta \, \exp [-i m \tau]~,~~
 \Psi_{\4} = \eta \, j \, \exp [-i m \tau]~.  
\]
The $i$-complex projection of inner products guarantees 
the orthogonality of the 
solutions of the complexified quaternionic Dirac equation.
For example, we get 
\[   \left\{  \overline{\Psi}_{\1}  \,  \Psi_{\2}  \right\}_{\mathbb{C}} =   
\left\{ \overline{\exp [i m \tau]}  \, \, j \,  
\exp [i m \tau] \right\}_{\mathbb{C}} =
\left\{ j \, 
\exp [2 i m \tau] \right\}_{\mathbb{C}}  
= 0~.\]
Many articles in the quaternionic literature use complexified
quaternions to formulate Lorentz transformations and 
Dirac equation, see for example~\cite{DeRod,MOR,LAN,CON,GUR2,MPL}
and references therein. 
In the paper of ref.~\cite{MOR}, the Lorentz group,
realized by a subset of $\mathsf{SL}(2,\mathbb{H})$, is then 
{\em diagonalized} by introducing an imaginary unity $\eta$
which commutes with the set $\left\{ i,j,k \right\}$.
In this paper, we have shown the possibility to formulate a quaternionic 
{\em one-dimensional} Lorentz group representation by using linear 
functions of real quaternions. The absence of $\eta$ is replaced by 
the use of the right-acting imaginary unit $R_{\i}$. Our quaternionic 
Lorentz group is isomorphic to 
$\mathsf{SL}(1,\mathbb{H}^{\L}\otimes \mathbb{C}^{\R})$.
By using the transformation properties of quaternionic 
spinors under space-time transformations we have obtained
a {\em new} quaternionic representation from the gamma matrices.
This representation, quaternionic counterpart of the standard chiral
representation, allows to rewrite the fermion sector of electroweak 
Lagrangian in terms of the space-time differential operators 
$\mathcal{D}_{\tpm}$. We have also showed that the Dirac, Majorana and 
chiral representations  for the gamma matrices diagonalize  respectively 
$\Upsilon^{\0}$,   $\Upsilon^{\1 \2 \3}$  and  $\Upsilon^{\5}$, 
anti-hermitian generators of the right-acting electroweak gauge 
group.

A quaternionic formulation of the Dirac equation (\ref{dirac})  
by linear function of real quaternions is found in the papers of 
Morita~\cite{MOR} and  Rotelli~\cite{ROT}.  We recall that without 
the introduction of the complex projection  of the inner product we cannot 
identify $\partial^{\mu} R_{\i}$ as momentum operator and cannot recover
four orthogonal solutions. The use of $i$-complex projection of 
quaternionic inner products is explicitly assumed in the Rotelli 
paper~\cite{ROT}.  So, the Rotelli work plays a  
{\em fundamental} role in the formulation  of a quaternionic quantum  
mechanics based on {\em complex inner products}. In such a formulation, 
the imaginary complex unit of the standard 
quantum mechanics is ``translated''  by the quaternionic imaginary unit $i$ 
in the quaternionic wave functions and by the right-acting imaginary unit 
$R_{\i}$ in quaternionic operators~\cite{DeRod}. Obviously, such an 
interpretation also extends to
quaternionic quantum fields, where the commutation relations are now given 
in terms of the right-acting opertor $R_{\i}$~\cite{SCA}.
Finally, the introduction of complex inner products is also important in 
discussing quaternionic tensor products~\cite{TEN}, 
group representations~\cite{DeDu} and gauge models~\cite{GAU}.

The use of complex projection of inner products,  within formulations 
of physical theories based on non-commutative algebras (quaternions,
complexified quaternions, STA, etc.), could play an 
important role in suggesting  unification group (see the discussion
on the electroweak model given in the previous section
and ref.~\cite{HES})  and searching for 
{\em hidden} geometric structures in relativistic wave equations.
The last point represents the main motivation of the Hestenes 
work~\cite{DH1,DH2,DH3}. The correct geometric interpretation of
the complex imaginary unit of the standard quantum mechanics still 
represents an open question in Clifford algebra. In fact,  
the complexified quaternionic 
Dirac equation~\cite{CON,GUR2,MPL} can also be written as 
\begin{equation}
\label{ps}
 \eta \, \left( \partial_{t} \Psi \, j  + 
\boldsymbol{h} \cdot \boldsymbol{\partial} \Psi \, k
\right) = m \, \Psi~.
\end{equation}
In this formulation, the complex imaginary unit  of standard quantum 
mechanics is identified with the imaginary unit $\eta$ of the complexified 
quaternionic field, $L_{\e} \Psi \equiv R_{\e} \Psi = \eta \Psi$ . 
In terms of STA, $\eta$ 
represents a pseudo-scalar.  We have ``two'' possible geometric 
interpretations of the complex imaginary unit of standard quantum 
mechanics, namely {\em bivector} if we select  $i$-complex projections   
and {\em pseudo-scalar} if we choose $\eta$-complex projections of
inner products.
 
In this paper, we have investigated a quaternionic formulation of
Lorentz transformations and Dirac equation by means of linear
functions of real quaternions. The main point is not the use of
{\em real} against {\em complexified} quaternions but the use of 
a complex projection of quaternionic  inner  products. Complex projections 
represent the common link for  many formulations of physical theories  
based on non-commutative algebras.

A different approach to quaternionic quantum mechanics, 
based on  {\em quaternionic} inner products is extensively developed 
in the literature, see for example the discussion on  
foundations of  quantum mechanics~\cite{FIN1,FIN2,FIN3}, 
CP-violation~\cite{AdlCP,DAV,AdlWu}, quantum field 
dynamics~\cite{AdlNP,ADLb} and preonic models~\cite{PRE}.

\acknowledgement{{\bf Acknowledgements.} The author thanks the
Department of Physics, University of Lecce, for invitation, 
hospitality and financial support. The author is greatly 
indebted to Prof. P.~Rotelli for helpful comments.
He also wishes to express his thanks to G.~Ducati and  C.~Nishi
for many stimulating conversations. Finally, the author thanks Prof. 
A.~A.~Gsponer for a private communication.  His suggestions and 
remarks helped to improve the notation and the presentation of this paper. 
}



\end{document}